\input{aipcheck}

\documentclass[
    ,final            % use final for the camera ready runs
  ]
  {aipproc}

\layoutstyle{6x9}
\usepackage{journal_shortcuts}

\begin{document}

\title 
      [Evolution of X-ray Cavities in Galaxy Clusters]
      {Evolution of X-ray Cavities in Galaxy Clusters}

\author{Marcus Br\"uggen}{
  address={Jacobs University Bremen, Campus Ring 1, 28759 Bremen, Germany},
}

\author{Evan Scannapieco}{
  address={School of Earth
and Space Exploration,  Arizona State University, P.O.  Box 871404,
Tempe, AZ, 85287-1404, USA},
  }
  
  \author{Sebastian Heinz}{
  address={Department of Astronomy, University of Wisconsin, 475 N Charter Street 
Madison, WI 53706, USA},
}


\begin{abstract}

The physics of X-ray cavities in galaxy clusters can be constrained by their
observed morphological evolution, which is dependent on such
poorly-understood properties as the turbulent density field and
magnetic fields. Here we combine numerical simulations that
include subgrid turbulence and software that produces synthetic X-ray
observations to examine  the evolution of X-ray cavities in the the
absence of magnetic fields. Our results reveal an anisotropic size
evolution that is very different from simplified,
analytical  predictions.  These differences highlight some of the key
issues that must be accurately quantified when studying AGN-driven
cavities, and help to explain why the inferred $pV$ energy  in these
regions appears to be correlated with their distance from the cluster
center.  Interpreting X-ray observations  will require detailed
modeling of effects including mass-entrainment, distortion by drag
forces, and  projection. Current limitations do not allow a
discrimination between purely hydrodynamic and magnetically-dominated
models for X-ray cavities.

\end{abstract}

\maketitle

%%%%%%%%%%%%%%%%%%%%%%%%%%%%%%%%%%%%%%%%%%%%
%% MAINMATTER
%%%%%%%%%%%%%%%%%%%%%%%%%%%%%%%%%%%%%%%%%%%%

\section{Introduction}

The nature of AGN-driven X-ray cavities remains one of the major
outstanding  questions in understanding the physics of cool-core
galaxy  clusters. While the  observed morphologies and sizes of the
cavities provide us with useful clues as to the processes at work in
these regions, interpretation of the observations is far from
straightforward.
Notably, it is unclear how far AGN-driven cavities rise in
the cluster, how they couple to the surrounding medium, and how they
evolve.

More specifically,  the presence of these cavities 
has raised a number of key questions.  The buoyant bubbles inflated by
the central AGN are unstable  to the Rayleigh-Taylor (RT) instability,
which occurs whenever a fluid  is accelerated or supported against
gravity by a fluid of lower density. Yet, these cavities appear to be
intact even after inferred ages of several $10^8$ yrs, as is the case for
the outer cavities in Perseus \citep{nulsen:05}. On the other hand, purely hydrodynamic
simulations fail to reproduce these observations.  Instead, the RT and other
instabilities shred the bubbles in a relatively short time
\citep{brueggen05, heinz:06}. The time scales of the RT in bubbles are calculated in \cite{pizzolato:06}. Magnetic fields have been
shown to alleviate this problem somewhat \citep{robinson:04, jones:05,
ruszkowski:07}, but they also reduce the extent to which the interior
of the hot bubbles couples to the surrounding medium, making it much
more difficult for AGN heating to balance cooling.

\begin{figure*}
\includegraphics[width=0.9\textwidth]{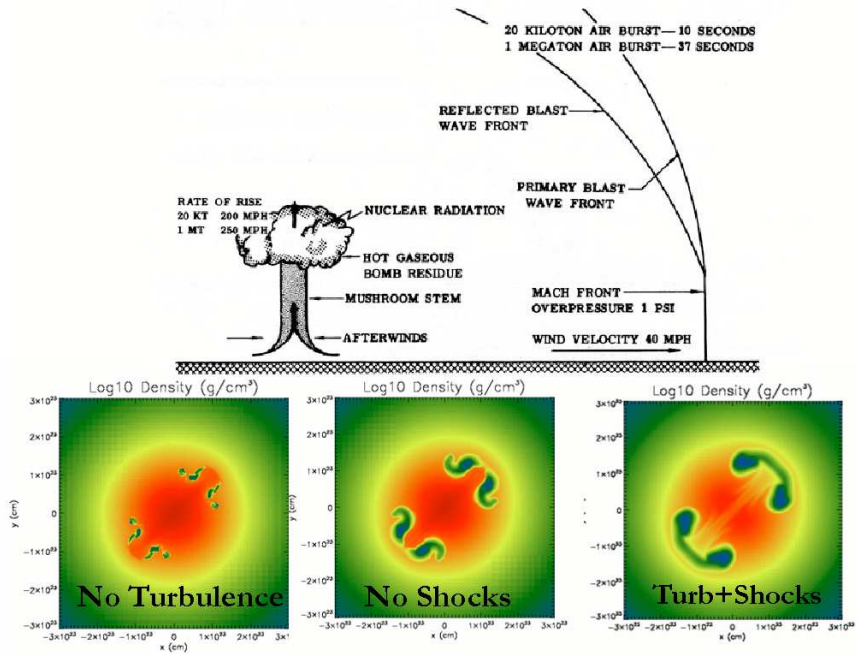}
\caption{Comparison of simulations with and without subgrid model that yields correct results for the Rayleigh-Taylor and Richtmyer-Meshkov instabilities. The panels show slices of the density. The left panel shows a simulation without a subgrid model for turbulence. The right panel shows the effect of supersonic bubble inflation (with a subgrid model).}
\label{fig:subgrid}
\end{figure*}

In a recent paper, Scannapieco \& Br\"uggen (2008) \cite{scannapieco:08} showed that
although pure-hydro simulations indicate that AGN bubbles are
disrupted into pockets of underdense gas, more detailed modeling of
turbulence indicates that this is a poor approximation to a cascade of
structures that continues far below current resolution limits.  Using
a subgrid turbulence model developed by Dimonte \& Tipton (2006), they
carried out a series of simulations of AGN heating in a cool-core
cluster with the adaptive mesh refinement code, FLASH.  These
simulations showed that Rayleigh-Taylor instabilities act on subgrid
scales to effectively mix the heated AGN bubbles with the ICM, while
at the same time preserving them as coherent structures (see Fig. \ref{fig:subgrid}). The AGN
bubbles are thus transformed into hot clouds of mixed material as they
move outwards in the hydrostatic medium, much as large airbursts lead to
a distinctive ``mushroom cloud'' structure as they rise in the
hydrostatic atmosphere of Earth. This allows X-ray cavities to remain
intact out to large distances from the cluster centre while still
coupling  to the surrounding medium.

%FFFFFFFFFFF
%FFFFFFFFFFF

Alternatively, it has been suggested that instead of underdense
pockets of ideal gas, the cavities are produced by magnetically
dominated jets \citep{li:06}. In 3D MHD simulations by
Nakamura et al. \cite{nakamura:06, nakamura:07} bubbles are inflated by a
current-carrying jet that injects magnetic flux into a small volume in
the vicinity of the supermassive black hole.  The jets are launched by
injecting non-force-free poloidal and toroidal magnetic fields, and
form large currents which travel along the inner jet axis into the
lobes and return on the outer boundary of the lobes, forming a sheath
around the jet axis.  Like bubbles of hot, underdense gas, such jets
expand  subsonically into wide lobes that appear cooler than the surrounding medium.
Unlike hot, underdense gas, these magnetically dominated bubbles behave differently as
they rise through the cluster atmosphere.

Recently, it was investigated whether one can use the
measured sizes of X-ray cavities observed at different locations in
their host clusters to discriminate between these two models \cite{diehl:08}.
Compiling the sizes and radial offsets of 64 cavities in the X-ray
halos of clusters and groups, they were able to show a tight
correlation between these two quantities, which is substantially
different than one would expect from simple analytic estimates in the
pure-hydro case.  From this comparison, they  came to the preliminary
conclusion that the data favor the current-dominated
magneto-hydrodynamic jet model.

While a useful first step, such analytical prescriptions neglect some
important physical effects that are likely to affect the
interpretation of the data. Whereas the loss of pressure due 
to bremsstrahlung radiation is negligible, mass entrainment produced by
hydrodynamic instabilities is likely to be important
\citep{pavlovski:07}. This will add to the growth of the
bubbles produced by the expansion during the ascent in the stratified
atmosphere. In fact, there is some circumstantial evidence for mass entrainment
in FR I radio sources.  Croston et al. (2008) \cite{croston:08} showed that equipartition
internal pressures are typically lower than the external pressures
acting on the radio lobes, so that additional non-radiating particles
must be present. A correlation between the structure of the radio
sources and the apparent pressure imbalance can be taken as
observational evidence that entrainment may provide this missing
pressure.  Moreover, drag forces and acceleration by buoyancy 
distort the bubbles, complicating the interpretation of bubble radii.

%%%%%%%%%%%%%%%%%%%%%%%%%%%%%%%%%%%%%%%%%%%%
%% Sample figure:
%%
%% The option [height=...] scales the picture to the given height,
%% without it it would be printed at its nominal size
%%%%%%%%%%%%%%%%%%%%%%%%%%%%%%%%%%%%%%%%%%%%

%*************
\section{Method}
%*************

\subsection{Code} \label{sec:code}

All simulations were performed with  FLASH version 3.0, a multidimensional adaptive mesh refinement hydrodynamics
code, which  solves the Riemann problem on a Cartesian grid using a
directionally-split  Piecewise-Parabolic Method (PPM) solver.  While
the direct simulation of turbulence is extremely challenging,
computationally expensive, and dependent on resolution  
\citep[e.g.,][]{glimm:01}, its behavior can be approximated to a good degree of
accuracy by adopting a subgrid approach. The details of the simulation are given in \cite{bruggen:09}. The reader is referred to this work for these and other related  discussions.

X-ray cavities in the ICM are thought to be inflated by a pair of ambipolar
jets from an AGN in the central galaxy that inject energy into small
regions at their terminal points, which expand until they reach
pressure equilibrium with the surrounding ICM. The result is a pair of underdense, hot bubbles
on opposite sides of the cluster centre. In order to produce bubbles, we started the simulation by injecting
a total energy of $E_{\rm inj}$
into two small spheres of radius $r_{\rm inj}$ = 4.5 kpc at distances of 13 kpc from
the cluster centre.  

To allow a direct comparison of
simulation output with X-ray data, we made use of a newly-developed
pipeline for post-processing of gridded simulation output.
The X-ray-imaging pipeline {\tt XIM} \citep[see also][]{heinz:09} is
a publically-available set of scripts that automate the creation of
simulated X-ray data for a range of satellites.  It takes as input the
density, temperature, and velocity, as well as a large number of
parameters and provides simulated X-ray data in the form of spectral-imaging 
data cubes.

{\tt XIM} is focused on visualizing X-ray data from thermal plasmas.
The scripts allow the choice of a user-supplied spectral model as well
as the default thermal {\tt APEC} plasma emission model
\citep{smith:01}, which self-consistently calculates the equilibrium
ionization balance for a thermal plasma.  For a fixed set of
abundances and a given temperature, {\tt APEC} then computes
interpolated high resolution model X-ray spectra.  Atomic data are
taken from the {\tt ATOMDB} using {\tt APED} \citep{smith:01b} and
combined with bremsstrahlung continuum for all species.

Having computed the spectral contribution in each cell,
{\tt XIM} then calculated a raw spectral data cube by projecting the
data along one of the three Cartesian coordinate axes.  Spectra were
emission-measure weighted and Doppler shifted with the user-provided
radial velocity, neglecting relativistic effects.  A user-supplied
tracer grid was used to weigh the data by the thermal plasma content
of the cell.  The spectral grid was oversampled by a factor of three with respect to
the final output energy grid to allow accurate representation
of Doppler shifts in the output spectra.

The data was further redshifted according to the user-specified
cosmological redshift and the coordinate axes are scaled to the proper
angular size, given the redshift and cosmological parameters. 
The flux in a cube (x,y,wavelength) was then scaled to the cosmologically correct 
flux at the given distance.
The projected data were processed for foreground photo-electric
absorption using the Wisconsin Absorption Model \citep[WABS][]{morrison:83}.

Next, the raw spectral-imaging cube (x,y,wavelength) was re-gridded in the
two spatial directions onto the detector plate scale of the
user-specified instrument.  {\tt XIM} incorporates telescope
parameters for  {\sc Chandra, Constellation-X, XEUS}, and  {\sc
XMM-Newton} and will incorporate a telescope model for IXO once
response matrices become available, 
in our case the Advanced 
Imaging Spectrometer (ACIS), on board the {\sc Chandra} X-ray observatory.
The re-gridded data cube was then
convolved with the appropriate spectral response and ancillary response matrices,
and the convolution output was regridded
onto the user-specified energy grid.

The output was convolved with a model point-spread function for the selected telescope.
The current version of {\tt XIM} is limited to a Gaussian point spread function with
energy-independent kernel width. It takes into account quantum efficiency and
telescope effective area, but neglects detector
non-uniformity, vignetting, and point-spread function variance. Finally, Poisson-distributed
photon counts were calculated for a user-specified exposure time.

\section{Results}

The differences produced by the subgrid turbulence are
significant, especially at times $\geq$ 150 Myrs. Without the subgrid
model, the bubble does not form a single coherent structure but rather looks
patchy, eventually coming apart into resolution-dependent subclumps
as described in Scannapieco \& Br\" uggen (2008).  The cavities in the run without 
subgrid turbulence also show a weaker X-ray surface brightness contrast,
even though the ambient material gets mixed into the bubble by subgrid
turbulence. Current observations of
X-ray cavities support this picture, and further observations of bubbles
at larger distances from cluster centres will help to probe the
stability of the bubbles and check the predictions from these kinds of
simulations.

\begin{figure*}
\includegraphics[trim=0 0 0 0,clip,width=0.45\textwidth]{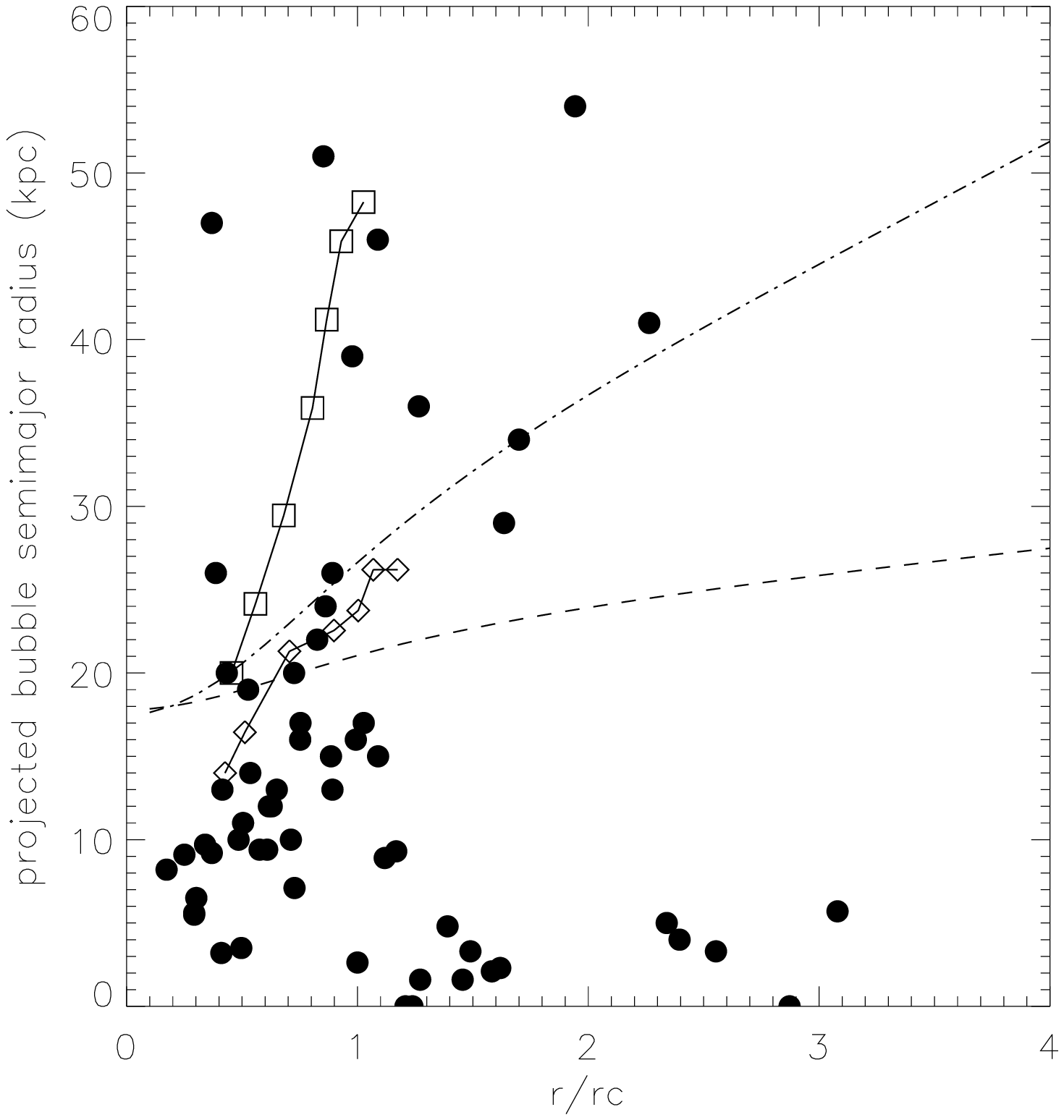} 
\includegraphics[trim=0 0 0 0,clip,width=0.45\textwidth]{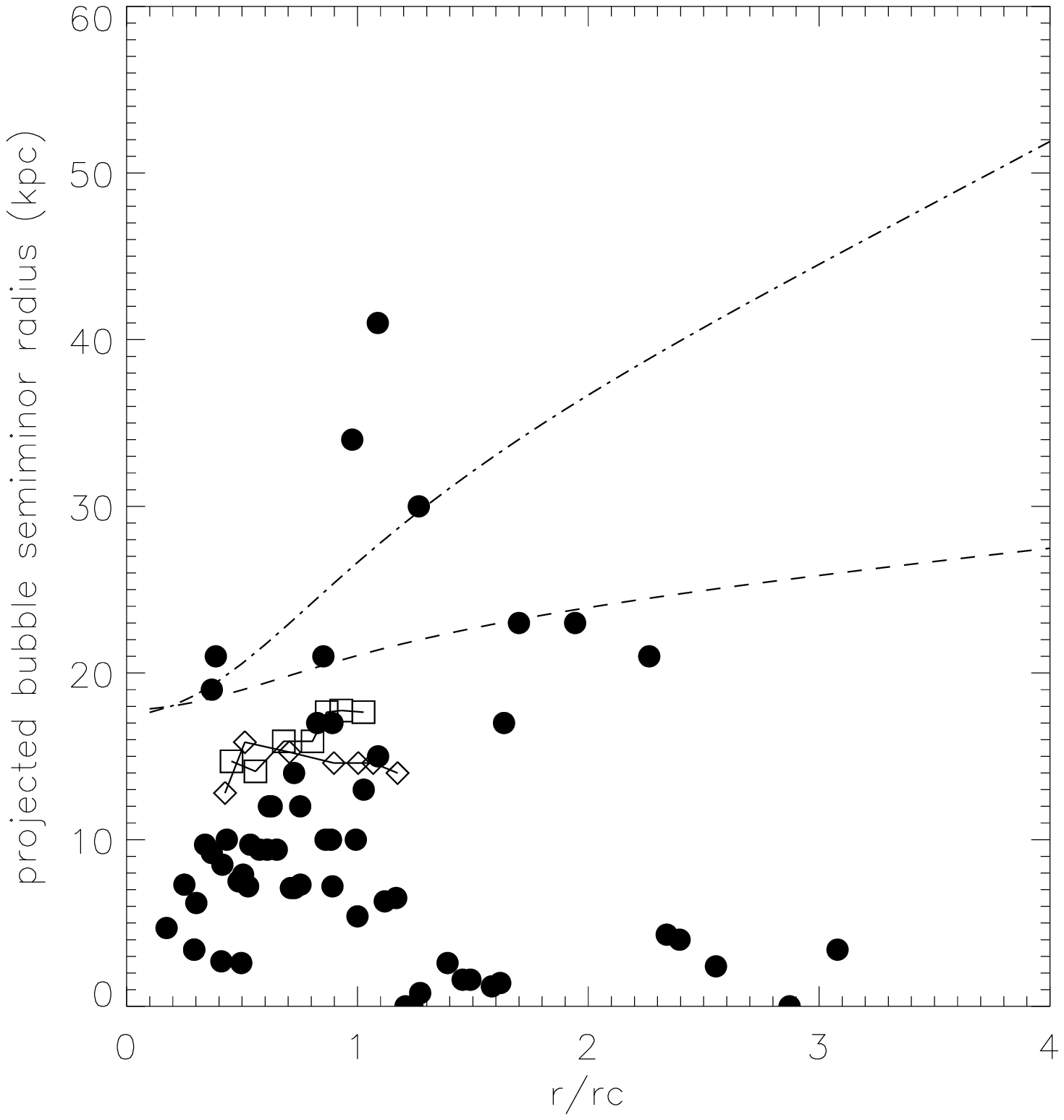} 

\caption{Plot of cavity projected radii versus their projected distance from the cluster centres. The connected squares correspond to the measured radii from synthetic observations where the axis of the AGN lies in the plane of the sky. The diamonds correspond to the case where the AGN axis is inclined by 45 degrees with respect to the line of sight. The unconnected filled circles give the sizes of observed cavities compiled in Rafferty et al. (2006). The dashed line shows the prediction for a spherical, adiabatically expanding bubble for $\Gamma=5/3$, and the dot-dashed line shows the prediction for a current-dominated jet. {\bf Left:} semimajor radii, {\bf Right:} semiminor radii.}
\label{fig:linradius}
\end{figure*}

The turbulence also affects the inferred bubble sizes, which have
been determined by subtracting a smooth, radially-symmetric background model from the
X-ray maps, as is often done with real observations.  The corresponding radii
are shown in Fig.~\ref{fig:linradius}. 

We find that analytic estimates as given by Diehlt et al. (2008) \cite{diehl:08} are not able to capture
effects such as mass-entrainment and distortion of the bubbles by drag
forces, which are naturally included in the simulation.  In fact, our
simulations show clearly that the cavities evolve aspherically as they
rise through the cluster, expanding quickly in the perpendicular
direction, but expanding slowly, or even becoming compressed, in the
radial direction.

In our hydrodynamic model of the cavity evolution, the effects of
projection are not trivial, and the data points are not just
systematically shifted toward smaller radii. The slopes of the
size-distance relationship of the bubbles become significantly
shallower as the AGN axis tilts toward the line of sight. The
appearance of the bubbles show stark differences, especially when the
bubbles are further than two bubble radii from the centre of the
cluster. Fig.~\ref{fig:linradius} suggest that most observed bubbles
come from systems where the AGN axis lies between 90 and 45 degrees
from the line of sight.  In fact, this is expected geometrically
for a random distribution of bubble axes.

In the case where the AGN axis lies at 45 degrees from the line of
sight, the bubbles look more like horseshoes. A confirmation of this
will require looking for bubbles at greater distances from the
centres in clusters where it is believed that the jet axis is inclined
not too far from the line of sight. This could help to verify whether
the bubbles are mainly hydrodynamic bubbles. One may speculate that
the giant cavities found in Abell 2204 resemble these late-stage
horse-shoe shaped bubbles \citep{sanders:08b}.

%FFFFFFFFFFF
%FFFFFFFFFFF

Finally, the $pV$ energy inferred from  the observations appears to increase with
increasing distance of the bubbles form the cluster centre. This may in part be due to the fact that bubbles tend to be overpressured close to their origin, and expand to reach
pressure equilibrium as the rise though the ICM. However, excess
pressure  in the observed cavities is difficult to measure, except
indirectly through the presence of shocks and sound waves.

A second possibility for the radial increase in inferred $pV$ is the
entrainment of ambient material.  In fact, this is the main reason for
the increase in our simulations, as in our model the bubbles quickly
reach  pressure equilibrium with the surrounding medium,  well before
moving noticeably out from the cluster centre.  Over the 200 Myrs that
we simulate, we find that the inferred $pV$ energy for the same bubble
grows by a factor of  3-5.  The measured energies are more
reliable when the bubble are farther from the centre of the cluster,
which, unfortunately, is where they are most difficult to observe.
Overall, the error introduced into estimates of the $pV$ energy by
these effects is of the same magnitude as the error related to the
unknown equation of state of the plasma inside the bubbles.

\section{Conclusions}

In this study we have focused on a model in which X-ray cavities
evolve purely hydrodynamically, and reconstructed their detailed
evolution using of two major tools:  AMR simulations that include
subgrid turbulence and synthetic X-ray software that produces realistic
observations from these simulations.  Together these tools allow us to
capture such important effects as mass-entrainment, distortion of the
bubbles by drag forces, and observational effects. These effects
lead to an evolution that is  drastically different than expected from
simple analytic estimates.

In particular, we find that while the radial extent of the cavities
changes slowly as  a function of distance and time, they expand
rapidly in the perpendicular  direction. The result is a complex
evolution that is highly dependent on viewing angle and difficult to
compare conclusively with observations.  Although analytic
estimates of non-magnetic models evolve too slowly to match the observations,
our simulations show that the evolution of the
semimajor and semiminor radii is not in obvious contradiction to the
data.  In fact, our simulations naturally reproduce the overall 
trend for inferred $pV$ energies of observed X-ray cavities to increase as 
a function of distance from the cluster centre, an effect that is largely
due to mixing of entrained material into the rising cavities.
The size evolution we find is in good agreement
with the measured bubble sizes, although there is a large spread in
observational data, and a strong dependence on the direction from
which the cavities are viewed.  Indeed, the flattening and projection
of X-ray cavities have much stronger effects on our results than
initial bubble size and radius.

\begin{theacknowledgments}

MB acknowledges the support by the DFG grant BR 2026/3 within the Priority
Programme ``Witnesses of Cosmic History'' and the supercomputing grants NIC
2195 and 2256 at the John-Neumann Institut at the Forschungszentrum J\"ulich.
All simulations were conducted on the Saguaro cluster operated by the 
Fulton School of Engineering at Arizona State University.
The results presented were produced using the FLASH code, a product of the DOE
ASC/Alliances-funded Center for Astrophysical Thermonuclear Flashes at the
University of Chicago. 

 \end{theacknowledgments}



\bibliographystyle{aipproc}   % if natbib is available

\bibliography{BIBLIOGRAPHY/icm_conditions,%
BIBLIOGRAPHY/bubble_evol%
}

\begin{thebibliography}{21}
\expandafter\ifx\csname natexlab\endcsname\relax\def\natexlab#1{#1}\fi
\providecommand{\enquote}[1]{``#1''}
\expandafter\ifx\csname url\endcsname\relax
  \def\url#1{\texttt{#1}}\fi
\expandafter\ifx\csname urlprefix\endcsname\relax\def\urlprefix{URL }\fi
\providecommand{\eprint}[2][]{\url{#2}}

\bibitem[{Nulsen} et~al.(2005)]{nulsen:05}
P.~E.~J. {Nulsen}, D.~C. {Hambrick}, B.~R. {McNamara}, D.~{Rafferty},
  L.~{Birzan}, M.~W. {Wise}, and L.~P. {David}, \emph{\apjl} \textbf{625},
  L9--L12 (2005), \eprint{arXiv:astro-ph/0504350}.

\bibitem[{Br{\"u}ggen} et~al.(2005)]{brueggen05}
M.~{Br{\"u}ggen}, M.~{Ruszkowski}, and E.~{Hallman}, \emph{\apj} \textbf{630},
  740--749 (2005), \eprint{arXiv:astro-ph/0501175}.

\bibitem[{Heinz} et~al.(2006)]{heinz:06}
S.~{Heinz}, M.~{Br{\"u}ggen}, A.~{Young}, and E.~{Levesque}, \emph{\mnras}
  \textbf{373}, L65--L69 (2006), \eprint{arXiv:astro-ph/0606664}.

\bibitem[{Pizzolato} and {Soker}(2006)]{pizzolato:06}
F.~{Pizzolato}, and N.~{Soker}, \emph{\mnras} \textbf{371}, 1835--1848 (2006),
  \eprint{arXiv:astro-ph/0605534}.

\bibitem[{Robinson} et~al.(2004)]{robinson:04}
K.~{Robinson}, L.~J. {Dursi}, P.~M. {Ricker}, R.~{Rosner}, A.~C. {Calder},
  M.~{Zingale}, J.~W. {Truran}, T.~{Linde}, A.~{Caceres}, B.~{Fryxell},
  K.~{Olson}, K.~{Riley}, A.~{Siegel}, and N.~{Vladimirova}, \emph{\apj}
  \textbf{601}, 621--643 (2004), \eprint{arXiv:astro-ph/0310517}.

\bibitem[{Jones} and {De Young}(2005)]{jones:05}
T.~W. {Jones}, and D.~S. {De Young}, \emph{\apj} \textbf{624}, 586--605 (2005),
  \eprint{astro-ph/0502146}.

\bibitem[{Ruszkowski} et~al.(2007)]{ruszkowski:07}
M.~{Ruszkowski}, T.~A. {En{\ss}lin}, M.~{Br{\"u}ggen}, S.~{Heinz}, and
  C.~{Pfrommer}, \emph{\mnras} \textbf{378}, 662--672 (2007),
  \eprint{arXiv:astro-ph/0703801}.

\bibitem[{Scannapieco} and {Br{\"u}ggen}(2008)]{scannapieco:08}
E.~{Scannapieco}, and M.~{Br{\"u}ggen}, \emph{\apj} \textbf{686}, 927--947
  (2008), \eprint{0806.3268}.

\bibitem[{Li} et~al.(2006)]{li:06}
H.~{Li}, G.~{Lapenta}, J.~M. {Finn}, S.~{Li}, and S.~A. {Colgate}, \emph{\apj}
  \textbf{643}, 92--100 (2006), \eprint{arXiv:astro-ph/0604469}.

\bibitem[{Nakamura} et~al.(2006)]{nakamura:06}
M.~{Nakamura}, H.~{Li}, and S.~{Li}, \emph{\apj} \textbf{652}, 1059--1067
  (2006), \eprint{arXiv:astro-ph/0608326}.

\bibitem[{Nakamura} et~al.(2007)]{nakamura:07}
M.~{Nakamura}, H.~{Li}, and S.~{Li}, \emph{\apj} \textbf{656}, 721--732 (2007),
  \eprint{arXiv:astro-ph/0609007}.

\bibitem[{Diehl} et~al.(2008)]{diehl:08}
S.~{Diehl}, H.~{Li}, C.~L. {Fryer}, and D.~{Rafferty}, \emph{\apj}
  \textbf{687}, 173--192 (2008), \eprint{0801.1825}.

\bibitem[{Pavlovski} et~al.(2007)]{pavlovski:07}
G.~{Pavlovski}, C.~R. {Kaiser}, and E.~C.~D. {Pope}, \emph{ArXiv e-prints}
  (2007), \eprint{0709.1796}.

\bibitem[{Croston} et~al.(2008)]{croston:08}
J.~H. {Croston}, M.~J. {Hardcastle}, M.~{Birkinshaw}, D.~M. {Worrall}, and
  R.~A. {Laing}, \emph{\mnras} \textbf{386}, 1709--1728 (2008),
  \eprint{0802.4297}.

\bibitem[Glimm et~al.(2001)]{glimm:01}
J.~Glimm, J.~W. Grove, X.~L. Li, W.~Oh, and D.~H. Sharp, \emph{J. Comput.
  Phys.} \textbf{169}, 652--677 (2001), ISSN 0021-9991.

\bibitem[{Br{\"u}ggen} et~al.(2009)]{bruggen:09}
M.~{Br{\"u}ggen}, E.~{Scannapieco}, and S.~{Heinz}, \emph{\mnras} \textbf{395},
  2210--2220 (2009), \eprint{0902.4242}.

\bibitem[{Heinz} and {Br{\"u}ggen}(2009)]{heinz:09}
S.~{Heinz}, and M.~{Br{\"u}ggen}, \emph{\apjs submitted}  (2009).

\bibitem[{Smith} et~al.(2001{\natexlab{a}})]{smith:01}
R.~K. {Smith}, N.~S. {Brickhouse}, D.~A. {Liedahl}, and J.~C. {Raymond},
  \emph{\apjl} \textbf{556}, L91--L95 (2001{\natexlab{a}}),
  \eprint{arXiv:astro-ph/0106478}.

\bibitem[{Smith} et~al.(2001{\natexlab{b}})]{smith:01b}
R.~K. {Smith}, N.~S. {Brickhouse}, D.~A. {Liedahl}, and J.~C. {Raymond},
  \enquote{{Standard Formats for Atomic Data: the APED},} in
  \emph{Spectroscopic Challenges of Photoionized Plasmas}, edited by
  G.~{Ferland}, and D.~W. {Savin}, 2001{\natexlab{b}}, vol. 247 of
  \emph{Astronomical Society of the Pacific Conference Series}, pp. 161--+.

\bibitem[{Morrison} and {McCammon}(1983)]{morrison:83}
R.~{Morrison}, and D.~{McCammon}, \emph{\apj} \textbf{270}, 119--122 (1983).

\bibitem[{Sanders} et~al.(2008)]{sanders:08b}
J.~S. {Sanders}, A.~C. {Fabian}, and G.~B. {Taylor}, \emph{ArXiv e-prints}
  (2008), \eprint{0811.0743}.

\end{thebibliography}

\IfFileExists{\jobname.bbl}{}
 {\typeout{}
  \typeout{******************************************}
  \typeout{** Please run "bibtex \jobname" to optain}
  \typeout{** the bibliography and then re-run LaTeX}
  \typeout{** twice to fix the references!}
  \typeout{******************************************}
  \typeout{}
 }

\end{document}